# Scanning tunneling spectroscopy of inhomogeneous electronic structure in monolayer and bilayer graphene on SiC

Victor W. Brar[1], Yuanbo Zhang[1], Yossi Yayon[1], Aaron Bostwick[2], Taisuke Ohta[2,3], Jessica L. McChesney[2,4], Karsten Horn[3], Eli Rotenberg[2], Michael F. Crommie[1]

[1]Department of Physics, University of California at Berkeley, Berkeley, California, USA and Materials Sciences Division, Lawrence Berkeley Laboratory, Berkeley, California, USA

[2]Advanced Light Source, Lawrence Berkeley National Laboratory, Berkeley, California, USA

[3]Fritz-Haber-Institut der Max-Planck-Gesellschaft, Berlin, Germany

[4]Montana State University, Bozeman, Montana, USA

We present a scanning tunneling spectroscopy (STS) study of the local electronic structure of single and bilayer graphene grown epitaxially on a SiC(0001) surface. Low voltage topographic images reveal fine, atomic-scale carbon networks, whereas higher bias images are dominated by emergent spatially inhomogeneous large-scale structure similar to a carbon-rich reconstruction of SiC(0001). STS spectroscopy shows a ~100meV gap-like feature around zero bias for both monolayer and bilayer graphene/SiC, as well as significant spatial inhomogeneity in electronic structure above the gap edge. Nanoscale structure at the SiC/graphene interface is seen to correlate with observed electronic spatial inhomogeneity. These results are important for potential devices involving electronic transport or tunneling in graphene/SiC.

The recent discovery of novel electronic properties in mechanically exfoliated graphene sheets [1, 2] has ignited intense exploration into this new two-dimensional material. This has led to a resurgence of interest in graphite grown epitaxially by heating SiC [3-7], which may provide opportunities for large scale integration of graphene in future nanoelectronics. One advantage of graphene/SiC is that the thickness of graphite grown on SiC can be precisely controlled to be either single or multiply layered depending on growth parameters [4, 6]. Electrons in single layer graphene/SiC are Dirac Fermions (just as for the mechanically exfoliated graphene samples), a fact evidenced by transport as well as angle-resolved photoemission (ARPES) measurements [6, 8]. One major difference, however, between graphene/SiC and exfoliated graphene samples is that graphene/SiC is grown on a complex SiC precursor layer whose electronic structure is not well understood [3, 9-18]. It is crucial for future applications involving patterning of graphene/SiC to understand the electronic properties of graphene sheets grown upon this SiC precursor interface layer. Scanning tunneling microscopy (STM) provides a very direct technique to probe the local electronic structure of this novel low-dimensional graphene system.

In this letter we present a scanning tunneling spectroscopy (STS) study of single and bilayer graphene grown epitaxially on a SiC(0001) surface. We observe dramatic variations in graphene/SiC microscopic topography when graphene is imaged at different bias voltages. Low voltage graphene imaging reveals fine, atomic-scale carbon networks, whereas higher bias images are dominated by emergent spatially inhomogeneous large-scale structure. Differential conductance (dI/dV) "point" spectroscopy reveals a robust electronic gap-like feature around zero bias for both monolayer and bilayer graphene/SiC, as well as significant spatial inhomogeneity at higher biases above the gap edge. These measurements highlight the importance of the underlying graphene/SiC interface layer in determining the electronic properties of graphene/SiC.

Our measurements were performed in an Omicron LT-STM at 4.8K and a base pressure of ~$10^{-11}$ torr. STM/STS measurements were made with chemically etched tungsten tips. STM tips were electronically calibrated for local spectroscopy by performing dI/dV measurements on clean Au(111) both before and after graphene/SiC measurement (this ensures that our tips were free of anomalies in their electronic structure since Au(111) is a well-known surface [19]). dI/dV curves were measured through lock-in detection of the ac tunneling current modulated by a 477 Hz, 5-10mV (rms) signal added to the junction bias (bias voltage is defined as the sample potential referenced to the tip). I(V) and dI/dV spectra were measured by fixing the tip position at one point and scanning the junction voltage under open loop conditions. Samples were prepared in UHV conditions and their thickness was calibrated using ARPES measurements, as described elsewhere[4]. Graphene/SiC samples were then transferred through air and placed in the STM chamber where they were annealed overnight at ~800° C in a UHV environment. After measurement by STM, the samples were again measured via ARPES and no discernible difference was seen in their electronic structure.

Fig. 1a shows a constant current STM image of the graphitized SiC(0001) surface, acquired with a -1.0 V tunneling bias. This image exhibits two distinct types of areas, labeled "1L" and "2L". Hexagonal reconstruction patterns with a periodicity of 17.8 ± 2 Å are observed in both the 1L and 2L regions (unit cell shown in Fig. 1a). This is attributed to a 6X6 reconstruction of the SiC precursor layer beneath the graphene [9, 10, 15, 20], and shows no sign of the long range buckling observed in suspended graphene flakes [21]. Although both 1L and 2L terraces display this reconstruction pattern, they also have marked differences. 1L terraces have a roughness of 0.2 Å (rms value at 1.0V over a 200Å x 200Å area) and display a fine honeycomb structure with a lattice constant of 2.4 ± 0.1Å (Fig. 1b), agreeing well with the expected 2.46Å lattice spacing of graphene. 2L terraces have a lower roughness of 0.1 Å and display a fine triangular grid having a lattice constant of 2.4 ± 0.1 Å (Fig. 1c). We can correlate the spatial frequency of 1L and 2L regions in our STM images with relative abundance of 1L and 2L regions as measured by ARPES, thus allowing us to identify 1L terraces as monolayer graphene and 2L terraces as bilayer graphene. The difference in observed atomic structure between 1L and 2L regions can be attributed to the Bernal stacking of two graphene sheets on top of one another in the bilayer case, which leads to the observation of a triangular lattice, as in the case of HOPG. This is also consistent with recent observations from other groups [20, 22].

The energy-dependence of the local density of states (LDOS) of monolayer and bilayer graphene can be seen in the spatially averaged dI/dV spectra of Fig. 2a. Surprisingly, both spectra show a "soft gap" of ~100meV centered at 0V (corresponding to $E_F$). STM images of graphene acquired at energies within this gap appear quite different from images acquired at energies outside of the gap range. Low-bias imaging (within the gap) of graphene monolayers (Fig. 2b) reveals the fine honeycomb pattern coexisting with the larger 6x6 reconstruction feature. As imaging bias is increased to -0.6V (Fig. 2c), however, the atomic structure of the graphene monolayer is no longer visible, and instead a disordered surface emerges out of the 6x6 reconstruction that is dominated by bright triangular trimer features ~10Å to a side. At even higher bias voltages these triangular features appear ordered in the same periodicity as the 6x6 reconstruction (see inset to Fig.2c). The bilayer case is slightly different. At low biases within the gap we observe the fine triangular mesh and larger scale 6x6 reconstruction pattern as before (Fig. 2d). At higher bias (Fig. 2e) disordered bright trimer features emerge out of the reconstruction as in the monolayer, but here the fine atomic-scale grid remains visible.

The disorder observed in graphene topographic images is reflected in the spatially inhomogenous electronic structure of the surface. This can be seen in Fig.3a which shows individual dI/dV point spectra taken at random points over a 60Å x 60Å patch of graphene monolayer. The spectra display huge variations (especially in the high energy regime away from the gap edges) that deviate strongly from the spatially averaged spectrum of Fig 2a. This strong heterogeneity also appears (although less pronounced) in a similar random sampling of point spectra measured over a bilayer graphene region (Fig. 3b). We are able to correlate some of the inhomogeneity observed in dI/dV spectra directly with disordered structure observed in topography. Fig. 3c, for example, shows a typical dI/dV spectrum measured at the site of a bright trimer feature in graphene monolayer (shown in the inset). This spectrum deviates strongly from the spatially averaged monolayer dI/dV (also plotted in Fig. 3c) and shows much higher LDOS in the energy range V < -0.2 volts [23].

Previous STM studies of SiC surface reconstructions have shown that, prior to graphene growth, a carbon-rich precursor layer forms [9, 10, 15, 20]. This layer is slightly disordered and is composed of trimer structures that arrange in a 6x6 surface reconstruction (relative to the 1x1 SiC(0001) unit cell) with a periodicity of $\sim 17 \pm 4$ Å. Previous STM images of this layer [9, 10, 15, 20] are very similar to our high bias images of graphene/SiC monolayer (Figs. 2c,d), and thus we believe that the larger scale features in our high bias images (and resulting graphene/SiC electronic heterogeneity) arise from SiC precursor structure.

A striking feature in our spectroscopic data is the gap-like feature at $E_F$ that is seen for both monolayer and bilayer graphene. The gap we observe is in contrast to recent ARPES results which show linear π-π* bands with a Dirac point (i.e., intersection) at 0.4eV below $E_F$ for monolayer graphene/SiC, and parabolic π-π* bands with a 100meV gap at the same energy for bilayer graphene/SiC [5, 8]. The energy gap observed in our STS data for monolayer graphene/SiC does not have the characteristics of a Dirac point, and so its cause is not obvious.

Previous STS studies of a related system, HOPG (graphite), show reduced LDOS at $E_F$, but typically do not show a gapped density of states[24-26]. One STS study of HOPG, however, has shown gap-like behavior that was interpreted as an electronic charging effect (i.e. a Coulomb blockade) [27, 28]. Low-energy spectroscopic structure found in other HOPG STS studies has been interpreted as due to phonons in the energy range 0-150 meV [29, 30]. One STS study of multilayer graphitized SiC samples shows a much wider gap-like feature than what is observed in our monolayer and bilayer samples [31].

We consider three possible explanations for the gap-like feature that we observe in the electronic structure of graphene/SiC: (i) electronic states in the underlying SiC layer, (ii) charging/band-bending effects, and (iii) inelastic coupling to surface excitations. Regarding the first possibility, it is not clear how a SiC precursor energy-gap could dominate the electronic structure of a metallic graphene sheet, and so we feel that this possibility is unlikely. The second possibility might arise due to intense electric fields near the STM tip ("band-bending"). This can lead to gating of graphene below the STM tip [32, 33] and resultant local accumulation of charge and altered electron tunneling probability. The third possibility arises from the opening of new electron tunneling channels due to inelastic electronic coupling to surface excitations (this can create steps in a dI/dV spectrum[34] that may give it a gap-like appearance). Previous measurements do show phonon and plasmon modes in HOPG that lie in the energy range of the gap-like structure that we observe for graphene [29, 30].

In conclusion, we have measured the local electronic structure of monolayer and bilayer graphene grown on SiC(0001). Our results indicate that the graphene electronic structure is strongly inhomogeneous and that the underlying precursor layer plays an important role in this behavior. We observe an unexpected ~100meV gap centered at $E_F$ in both the monolayer and bilayer graphene, and the origin of this feature is not clear. Regardless of its origin, however, the existence of this energy gap has important implications for future potential devices utilizing electronic transport or tunneling through graphene/SiC layers.

We thank D.-H. Lee and G. Rutter for helpful discussions. This work was supported in part by NSF Grant EIA-0205641 and by the Director, Office of Energy Research, Office of Basic Energy Science, Division of Material Sciences and Engineering, U.S. Department of Energy under contract No. DE-AC03-76SF0098. One of the authors (Y. Z.) acknowledges a fellowship from Miller Institute for Research in Basic Science, University of California at Berkeley.

**Figure captions:**

**Figure 1.** (a) Constant current STM image of graphene/SiC (-0.1V, 0.3nA). Monolayer and bilayer graphene regions are labeled "1L" and "2L" respectively.  6x6 reconstruction unit cell is drawn for both types of terraces. (b) and (c) constant current image showing atomic structure of monolayer and bilayer regions, respectively (-0.02V, 0.01 nA). Positions of carbon atoms are drawn.

**Figure 2.** (a) Spatially averaged dI/dV spectra of graphene monolayer and bilayer regions.  (b,d) Low bias images of monolayer and bilayer regions respectively (-0.05V, 0.025nA). (c,e) The same regions measured at higher bias (-0.5V, 0.025nA).  Inset: Image of monolayer region at an even higher bias (-1.0V, 0.003nA).

**Figure 3.**  (a) Individual dI/dV spectra measured at different points within a 60Å x 60Å area of a graphene monolayer surface.  (b) Individual dI/dV spectra measured at different points within a 60Å x 60Å area of graphene bilayer surface.  (c) Individual dI/dV spectrum (black line) taken at the site of the trimer feature shown in the inset (marked by "X") compared to the average dI/dV monolayer spectrum (dashed blue line). Inset: dI/dV map taken at -0.35V highlighting trimer features.

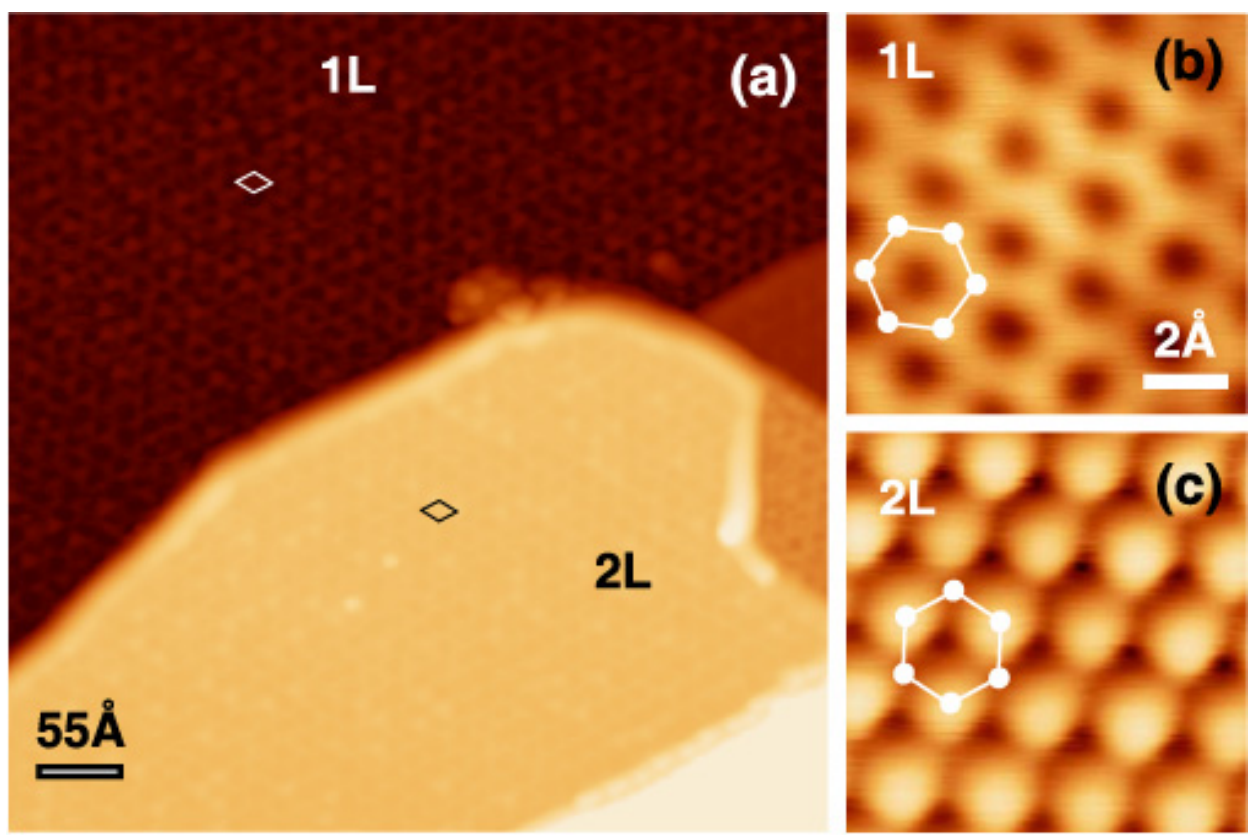

Fig. 1 V. W. Brar *et. al.*

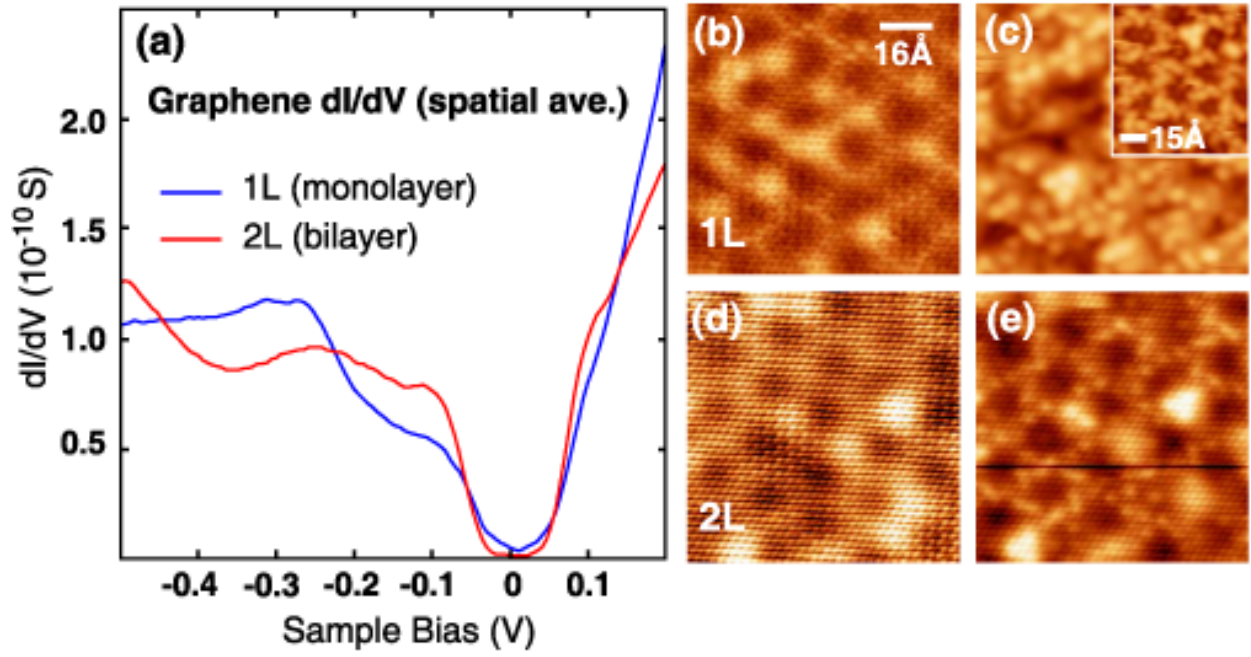

Fig. 2 V. W. Brar *et. al.*

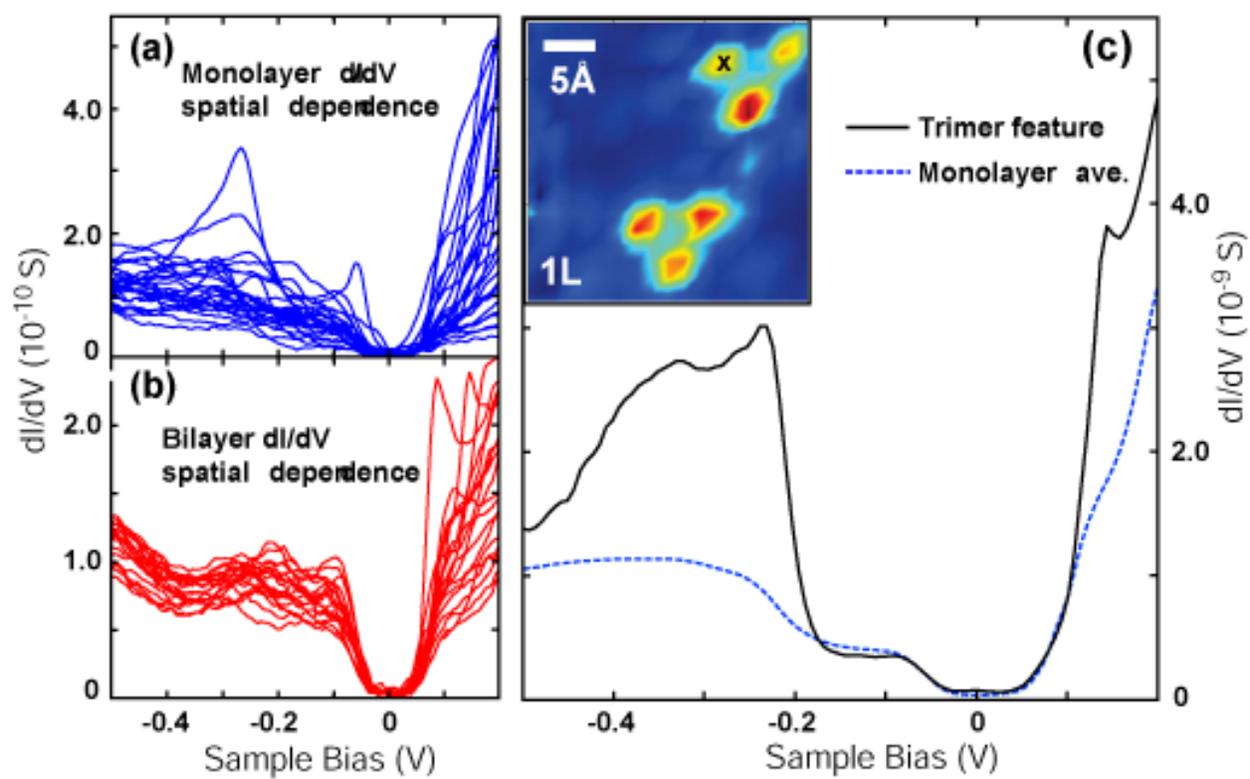

Fig. 3  V. W. Brar *et. al.*